\documentclass[prl,twocolumn,showpacs,amsmath,amssymb,aps]{revtex4}

\usepackage{graphicx}

\begin{document}

\title
{On the origin of the $\lambda$-transition in liquid Sulphur}

\author{T.~Scopigno$^1$, S.N.~Yannopoulos$^{2}$, F.~Scarponi$^{3}$, K.S. Andrikopoulos$^4$, D. Fioretto$^{3,1}$, G. Ruocco$^{5,1}$ }
\email{tullio.scopigno@roma1.infn.it}

\affiliation{ $^1$Research center SOFT-INFM-CNR,  Universit\`a di
Roma ``La Sapienza,'' I-00185, Roma, Italy} \affiliation{ $^2$
ICE/HT-FORTH GR-26504, Patras, Greece} \affiliation{ $^3$
Dipartimento di Fisica, Universit\'a di Perugia, via Pascoli,
I-06123 Perugia, Italy} \affiliation{$^4$ Physics Division, School
of Technology, Aristotle University of Thessaloniki, GR-54124,
Thessaloniki, Greece} \affiliation{$^5$ Dipartimento di Fisica,
Universit\'a di Roma "La Sapienza", 00185 Roma, Italy }

\begin{abstract}
Developing a novel experimental technique, we applied photon
correlation spectroscopy using infrared radiation in liquid
Sulphur around $T_\lambda$, i.e. in the temperature range where an
abrupt increase in viscosity by four orders of magnitude is
observed upon heating within few degrees. This allowed us -
overcoming photo-induced and absorption effects at visible
wavelengths - to reveal a chain relaxation process with
characteristic time in the ms range. These results do rehabilitate
the validity of the Maxwell relation in Sulphur from an apparent
failure, allowing rationalizing the mechanical and thermodynamic
behavior of this system within a viscoelastic scenario.
\end{abstract}
\date{\today}
 \maketitle

Common wisdom holds that liquids become less viscous as the
temperature is raised. On a microscopic ground, indeed, viscosity
can be regarded as arising from continual brushing of molecules
which are close to each other. On increasing temperatures the more
vigorous thermal motion of the molecules is expected to render
such mutual friction progressively less effective. This naive
description provides a qualitative interpretation of the decrease
in viscosity normally observed in almost all substances as the
temperature increases. There exist, however, a few remarkable
exceptions exhibiting complex behavior, which can be regarded as
extreme examples of how critically viscosity depends upon
molecular interactions. Specifically, there are ranges of
temperatures in which the viscosity of Sulphur exhibits anomalous
temperature dependence. In the liquid state, just above melting
($T_m$$=$$119$ $^o$C), eight-membered rings (S8) are the most
abundant species, and the viscosity slightly decreases with
temperature as normally expected, reaching values as low as ~0.01
Pa$\cdot$s at $T=157$ $^o$C. Further increase of temperature
causes a dramatic increase of viscosity \cite{bac_43}, accompanied
by gross changes in optical \cite{don_85,hos_94,sas_86} and
thermodynamic \cite{wes_59,kel_18} properties, which, at
$T$$\approx$$185$ $^o$C, reaches a maximum value of ~100
Pa$\cdot$s. Beyond this temperature, a gradual viscosity decrease
is observed. Although this phenomenon, known as
$\lambda$-transition ($T_{\lambda}=159$ $^o$C), has been tackled
for more than 150 years, its comprehension is still far from being
reached: to rationalize this puzzling temperature dependence of
viscosity, indeed, a liquid-liquid transition from a monomeric
(i.e. a S8 rings liquid) to a polymeric phase has been invoked
\cite{tob_59,gre_98}. The temperature dependence of the mass
fraction of S atoms participating in the polymeric component turns
out to be a central parameter for extracting information on many
thermodynamic aspects of the $\lambda$-transition
\cite{kal_03,koh_70}. This phenomenon has been classified as a
living polymerization \cite{gre_96} transition that essentially
involves two steps, the initiation (formation of diradicals
through opening of S8 rings) and propagation (concatenation of
species to form long polymeric chains). Thermo-chemical models
were developed in the past \cite{pow_43,tou_66,eis_69} to relate
the transport properties of liquid sulphur to the underlying
polymerization phenomenon. There are, however, provocative aspects
of the $\lambda$-transition still not fitting into the proposed
pictures. These are mostly related to the difficulty of casting
the mechanical properties of S within the usual viscoelastic
scenario,
up to the extent that a break down
of the Maxwell relation has been proposed for this liquid
\cite{koz_04}.

In this Letter, we report on the experimental evidence of a
previously unobserved low frequency relaxation, detected in the
1-10 kHz region by means of infrared photon correlation
spectroscopy (IR-PCS). The application of optical spectroscopy
techniques is critical in liquid Sulphur, since its absorption
spectrum is considerably different \cite{hos_94,sas_86} between
the monomeric and the polymer-dominated phase. Specifically,
undesired absorption and photo-induced changes occur when studying
the liquid with visible laser wavelengths on increasing
temperature above $T_\lambda$ \cite{sak_95}. To overcome these
problems, we developed a new technique employing a near infrared
radiation as probing field ($\lambda$=1064 nm), namely Infra-Red
Photon Correlation Spectroscopy (IRPCS). The revealed relaxation,
which implies the existence of a temperature dependent plateau
\cite{FERRY} of the elastic modulus, allowed us to give a coherent
picture of mechanical and thermodynamic properties of S within a
viscoelastic scenario through a modified version of the Maxwell
relation.

Infrared Photon Correlation Spectroscopy (IRPCS) was performed
using a solid state laser source operating at $\lambda=1064$ nm.
The detector was a Perkin Elmer avalanche photodiode retaining a
2$\%$ quantum efficiency at the probe wavelength, with a dark
count rate of ~50 counts/sec and an after pulse probability lower
than 0.3$\%$. The incident beam was focused onto the sample and
the scattered radiation was collected by a lens and then
collimated onto an optical fibre through an IR optimized
collimator from OZ-Optics. The scattering angle $\Theta=90^0$
defines the momentum transfer, Q, of the experiment through the
relation $Q=\frac{4\pi n}{\lambda}sin \frac{\Theta}{2}$, n being
the refractive index of Sulphur ($n^{1064}\simeq 1.847 (1.863)$ -
the first value refers to 140 $^o$C while the second one in
parenthesis to 183 $^o$C). The sample was heated in a furnace
whose temperature, measured by means of a platinum resistance
thermometer, was stabilized within 0.1 $^o$C.
\begin{figure}[h]
\includegraphics[width=.5\textwidth]{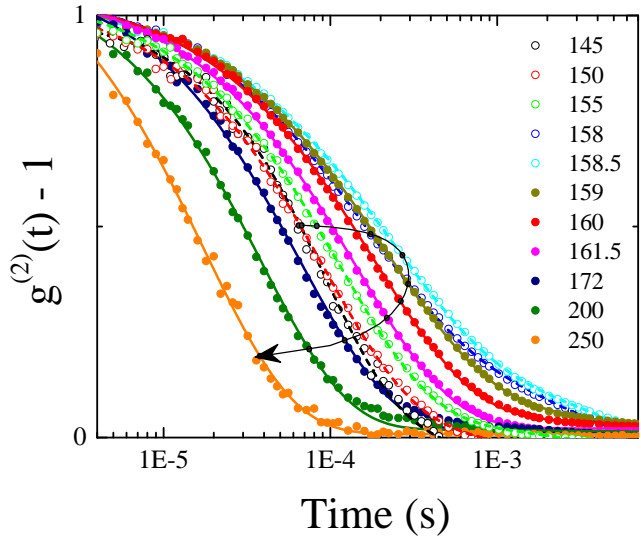}
\vspace{-5.5cm} \caption{Homodyne IR correlation functions at
selected temperatures ($^o$C) for $T< T_\lambda$ ($\circ$), and
$T> T_\lambda$ ($\bullet$). The T-dependence of the chain
relaxation is indicated by the arrow.}
 \label{f2}
\end{figure}
The digital signal coming out from the detector was acquired by a
National Instruments card and processed by a software package
(PhotonLab \footnote{PhotonLab is a Python extension for the
acquisition and analysis of photon counts developed by R. Di
Leonardo}) performing real time autocorrelation. Correlation
functions were collected at constant temperatures during an upscan
in the range $145<T<250$ $^o$C, with an integration time of 30 min
after 20 min stabilization time at each temperature. The
temperature step was half degree around the transition, and
coarser far from the transition.

\begin{figure}[h]
\includegraphics[width=.5\textwidth]{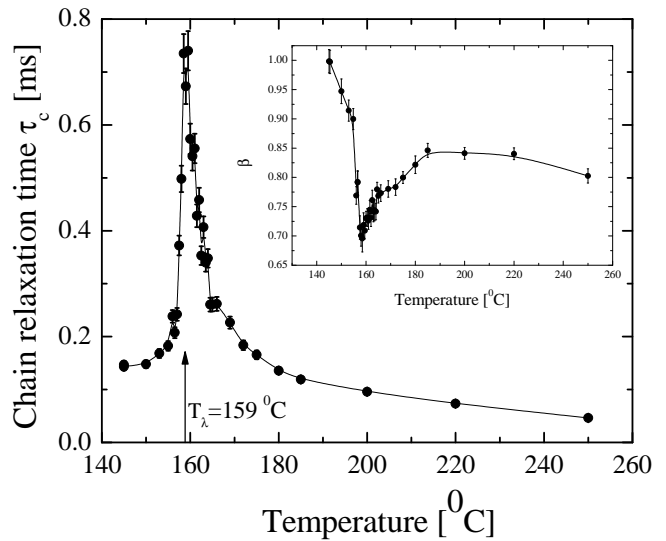}
\vspace{-5.5cm} \caption{Temperature behaviour of the chain
relaxation, as measured by IRPCS. The line is a guideline for the
eyes. Inset: T-dependence of the stretching parameter.}
 \label{f3}
\end{figure}

Selected examples of the autocorrelation functions of the
scattered intensity measured in in a broad temperature range
around the $\lambda$-transition are reported in Fig. \ref{f2}. The
existence of a previously unobserved relaxation process can be
easily observed and, already from the raw data, an unusual
behavior is detected: Upon heating above the melting point the
relaxation time is in the $10^{-4}$ s range, and it rapidly
increases reaching its longer timescale at $T_\lambda$, then it
decreases up to the higher measured temperature T=250 $^o$C. To
extract the relevant parameters, the reduced homodyne intensity
autocorrelation functions, the so called $g^{(2)}(t)$, have been
modelled by a stretched exponential function, representing the
decay displayed by the raw data, according to the equation:
\begin{equation}
g^{(2)}(t)-1=\phi (Q,t)=\left [e^ {\left (t / \tau \right
)^{\beta}} \right ]^2 \label{omo}
\end{equation}
where $\phi_Q(t)$ is the normalized density autocorrelation
function \footnote{The first equality in Eq.(\ref{omo}) holds in
the gaussian approximation, assuming an isotropic polarizability
\cite{BERNE}}. In fig. \ref{f3} we report both the average
relaxation time $\tau_c = \langle \tau
\rangle=\beta^{-1}\Gamma(\beta^{-1})\tau$ ($\Gamma$ is the Euler
gamma function), and the stretching parameter $\beta$ as a
function of $T$.
The temperature behaviour of $\tau_c$ clearly indicates its tight
relation with the $\lambda$-transition.

The discovery of this new relaxation process in the ms timescale
allows reconciling the unusual behaviour of sulphur with the usual
viscoelastic framework: in principle, according to one of the most
venerable phenomenological equations of physics, namely the
Maxwell relation $\eta=G_\infty \tau_\alpha$, the abrupt increase
in viscosity ($\eta$) observed at $T_\lambda$ should trigger an
equivalent increase of the structural relaxation time
$\tau_\alpha$ which should rise from ~1 ps to ~10 ns
\footnote{$G_\infty$$\approx$$10^8$$\div$$10^{9}$ Pa is the
"unrelaxed" shear modulus of a liquid, i.e. that measured in the
short wavelength limit}.
\begin{figure}[h]
\includegraphics[width=.5\textwidth]{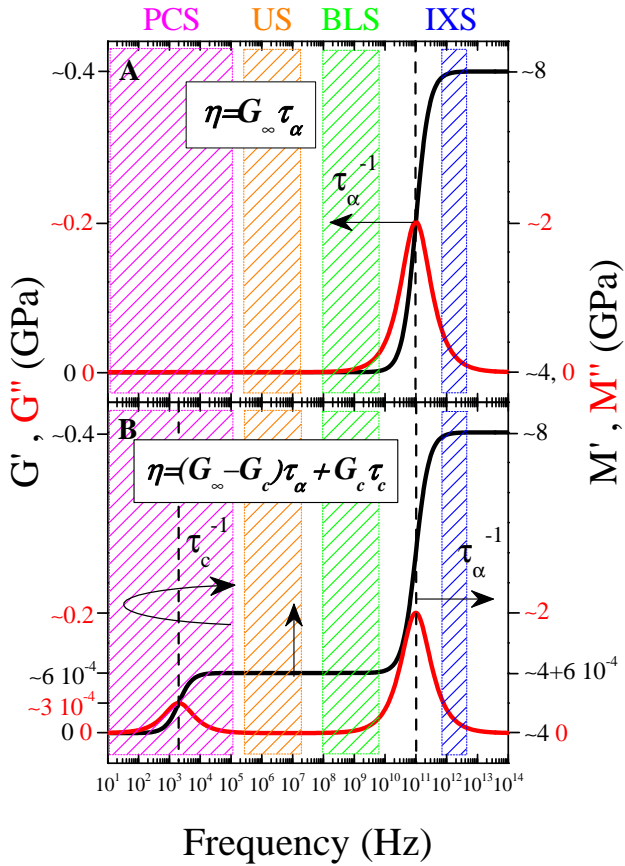} \vspace{-.8cm} \caption{The
Maxwell relation at work in liquid S. A: Sketch of the wrong
scenario which would lead into an apparent failure \cite{koz_04}
of the Maxwell equation. The increase of $\eta$ with temperature
should reflect an increase in $\tau_\alpha$
($\eta$$=$$G_\infty$$\tau_\alpha$), ultimately moving the
resonance condition ($\omega$$\tau_\alpha$$=$$1$) from the THz
(IXS) down to the GHz region (BLS). B: The working scenario: upon
heating, the $\alpha$ process shifts towards higher frequencies as
in any ordinary fluid (due to the decrease of $\tau_\alpha$),
while the rapid emergence of a secondary plateau at $T_\lambda$
(graphically emphasized for presentation reasons), observed in
$M'$ and $G'$ between 10 Khz and 10 Ghz, accounts for the unusual
viscosity increase observed in S.}
 \label{f1}
\end{figure}
This should imply the detection of the $\alpha$-relaxation in the
GHz domain by means of those techniques, like Brillouin light
scattering (BLS), that measure the longitudinal modulus
$M=K+\frac{4}{3}G$ (K being the bulk modulus). Indeed, when the
Brillouin peak position, $\omega$, satisfies the condition $\omega
\tau_{\alpha} =1$ relevant effects are expected on both sound
velocity and damping. More specifically, as sketched in Fig.
\ref{f1}A for a prototypical Debye-like relaxation, one should
have an increase of both the real ($M'$) and imaginary ($M''$)
parts of the longitudinal modulus, observed as an increase in the
measured sound velocity and acoustic absorption, respectively.
Interestingly, none of the two effects has ever been detected in
Brillouin light scattering experiments on Sulphur. On the
contrary, both $M'$ and $M''$ display a continuous smooth
decreasing trend across the $\lambda$-transition
\cite{sca_06,alv_96} testifying a decrease of the structural
relaxation time (the orthodox behaviour in ordinary liquids), and
not the increase expected on the basis of the Maxwell relation.
The situation becomes even more puzzling considering ultrasonic
experiments \cite{koz_04,tim_84,koz_01} performed in the 1-10 MHz
window, where the timescale of the probe is much longer than the
10 ns expected for the relaxation time. In this "relaxed", $\omega
\tau_\alpha <<1$, regime one should measure the viscosity directly
from the attenuation of an ultrasonic pulse. Surprisingly, no
critical temperature dependence of the acoustic damping has ever
been observed, as recently emphasized \cite{mon_05}. The attitude
of Sulphur of escaping for more than a century the established
scenario valid for any other molecular liquid, warranted the
labelling of a "viscous but non viscoelastic liquid"
\cite{koz_04}.

\begin{figure}[h]
\includegraphics[width=.5\textwidth]{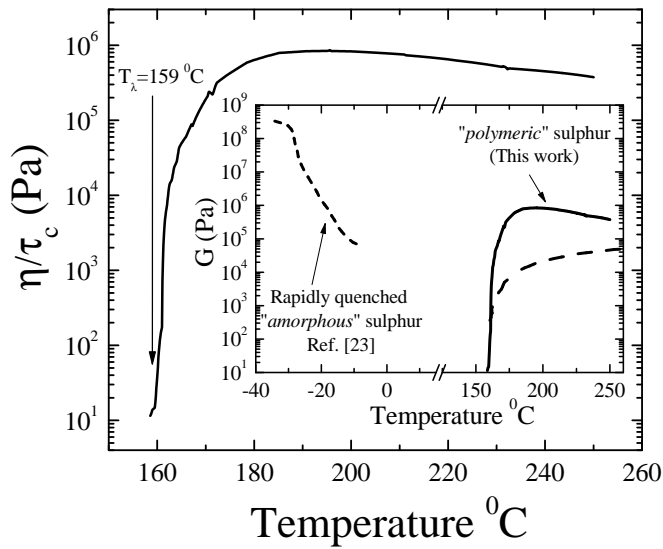}
\vspace{-5.5cm}\caption{Shear modulus $G_c$, as determined by the
ratio $\left [ \eta / \tau_c \right ]$, $\eta$ is from
\cite{bac_43}. Inset: shear modulus on an extended scale with low
temperature measurements in a quenched sulphur sample
\cite{tob_80} ($\cdot \cdot \cdot$) along with a theoretical
prediction \cite{eis_69} for the high temperature modulus ($---$).
}
 \label{f4}
\end{figure}

The discovery of the relaxation process reported here around and
above $T_\lambda$, allows to reformulate the Maxwell equation
within a two relaxation processes framework.
\begin{equation}
\eta=(G_\infty - G_c) \tau_\alpha + G_c \tau_c \label{maxwell_2}
\end{equation}
where $G_c$ is the shear modulus related to the new relaxation
process \footnote{The ground to Eq.(\ref{maxwell_2}) is provided
by the relation between the shear viscosity $\eta$ and the memory
function of the transverse current autocorrelation, $K_t(Q,t)$:
$\eta=\lim_{Q\rightarrow 0}\int_0^\infty
{K_t(Q,t)dt}=K_t(t=0)\tau$ in which the latter equality defines
the Maxwell relaxation time \cite{BY}. In presence of two
relaxation processes with significantly different timescales (as
in the present case) $K_t(Q,t)$ is the sum of two contributions,
as shown for instance in \cite{lev_2t}.}. Since $\tau_\alpha <
\tau_\alpha (T_m) \approx 10^{-11}$ s, $\tau_c >> \tau_\alpha$
holds, and being $G_\infty \approx 5 \cdot 10^8$ Pa \cite{tob_80}
the first term of (\ref{maxwell_2}) becomes progressively less
important as temperature is raised above $T_\lambda$. The increase
of viscosity with temperature, therefore, is triggered by the
appearance of the low frequency relaxation, which largely
compensates for the decrease of $\tau_\alpha$. Hence, one can
estimate the magnitude of $G_c$ as:
\begin{equation}
G_c=\left [ \frac{\eta - G_\infty \tau_\alpha}{\tau_c -
\tau_\alpha} \right]_{T>T_\lambda} \approx \left [
\frac{\eta}{\tau_c} \right ]_{T>T_\lambda}\label{GR}
\end{equation}
As can be seen from Fig. \ref{f4}, $G_c$ extracted through Eq.
(\ref{GR}) displays an abrupt temperature dependence growing by
five order of magnitude in ten degrees above the onset temperature
$T_\lambda$ and then asymptotically reaches values in the narrow
range $4\div 8 \cdot 10^5$ Pa. The presence of this additional
relaxation, leading to the reformulation of the Maxwell equation
given in Eq. (\ref{maxwell_2}), manifests itself through the
existence of this "intermediate" plateau in $G'$ (and consequently
$M'$), as sketched in Fig. \ref{f1}B. More importantly, at
variance with the $\alpha$-relaxation plateau $G_\infty$, the
plateau value $G_c$ turns out to have significant temperature
dependence around $T_\lambda$ and, in spite of its small value
($G_c$$\approx$$10^{-3}$$G_\infty$), the large contribution to
$\eta$ is brought about by the large value of $\tau_c$.
%
The discovery of this low frequency plateau above $T_\lambda$ also
allows to rationalize the abrupt decay of the shear modulus
measured by stress relaxation technique \cite{tob_80}, in the
range $-40<T<-30$ $^o$C, in sulphur rapidly quenched from T=250
$^o$C, and substantiates the hypothesis made in Ref.
\cite{mon_05}. Indeed, the strong increase of the plateau with
temperature just above $T_\lambda$ reported in the present study
can be traced back to the formation of polymeric chains of growing
average length \cite{pou_62,pou_63} (living polymerization
\cite{gre_96}). Upon an ideal, instantaneous quenching from the
polymeric phase, one can reasonably expect the polymer structure
and ultimately the average chain length to be frozen, i.e.
equivalent to those of the initial equilibrium state. In this
respect, we believe that the large difference $G_\infty-G_c$ that
we observe can be related with the decay of the shear modulus
reported in Ref. \cite{tob_80} : the value of the plateau in
$G_c(T)$ reported here  is $\approx 8\cdot 10^5$ Pa, while in the
decay observed in Ref. \cite{tob_80} the lowest measured value
upon heating is $\approx 6\cdot 10^4$ Pa. We consider such
difference as remarkably small, in view of the large variation of
the elastic modulus from the solid to the polymeric regime.
Moreover, this difference could be ascribed to the finite
quenching time which prevents the system from falling
instantaneously out of equilibrium. As a result there is a
reduction of the initial effective quenching temperature and,
correspondingly, the value of the elastic modulus at which the
structure is frozen is lower. Interestingly, the behaviour of the
high temperature modulus can be theoretically predicted
\cite{eis_69} on the basis of its measured low temperature value
\cite{tob_80} for an ideal quenching, i.e. assuming that the
average chain length is not affected by a rapid cooling. As can be
seen in Fig. \ref{f4}, the predicted $G_c(T)$ closely resembles
the shape of the plateau that we found.

In conclusion, in this Letter we present the experimental proof
for the existence of a KHz relaxation in liquid sulphur, providing
the rationale behind the $\lambda$-transition and ruling out a
simple coupling between viscosity and structural relaxation in
this system. At variance with ordinary liquids, where the increase
in viscosity is ruled by a direct proportionality (through the
constant $G_\infty$) with the structural relaxation time, in
liquid sulphur the large increase in viscosity reflects the large
temperature dependency of a relatively small intermediate plateau
in the real part of the shear modulus. According to the Maxwell
relation, the T-dependence of the viscosity is triggered and
amplified by a relatively large relaxation time. We speculate that
the nature of this plateau may be analogous to that usually found
in dense solutions of uncross-linked polymers (referred to as
"rubbery plateau" for its small value compared to the infinite
frequency one) and reflects the entanglement between the chains in
the melt \cite{FERRY,eis_69} which, in the case of the
$\lambda$-transition, proceeds first through a rapid increase of
the average chain length, thus originating the sharp onset of the
$G_c(T)$ plateau. We are grateful A. Sokolov for useful comments.


\end{document}